\documentclass[11pt,a4paper]{article}
\pdfoutput=1 
\usepackage[utf8]{inputenc}
\usepackage{amsmath}
\usepackage{amsfonts}
\usepackage{amssymb}
\usepackage{xcolor}
\usepackage{jheppub}
\usepackage{graphicx}

\title{Non-thermal production of Dark Matter after Inflation}
	 
\author[a]{Nicolás Bernal,}
\author[b]{Arindam Chatterjee,}
\author[b]{Arnab Paul.}

\affiliation[a]{Centro de Investigaciones, Universidad Antonio Nariño\\
Carrera 3 Este \# 47A-15, Bogotá, Colombia}
\affiliation[b]{Indian Statistical Institute\\
203 B.T. Road, Kolkata-700108, India}
\emailAdd{nicolas.bernal@uan.edu.co}
\emailAdd{arindam.chatterjee@gmail.com}
\emailAdd{arnabpaul9292@gmail.com}

\abstract{The existence of Dark Matter (DM) has been well established 
from various cosmological and astrophysical evidences. However, the particle 
properties of DM are largely undetermined and attempts to probe its 
interactions with the Standard Model (SM) particles have, so far, not 
met with any success. The stringent constraints on the DM-SM interactions, 
while does not exclude the standard lore of producing weakly massive interacting 
particle DM candidates through thermal freeze-out mechanism in its entirety, 
have certainly cast shadow on the same.
In this work, we consider non-thermal production of DM within a simple 
extension of the SM including an inflaton field and a scalar DM candidate. Assuming 
negligible interactions between the SM particles and the DM, we study the 
production of the latter at the end of inflation, during the (p)reheating 
epoch. In this context, we explore the role of DM self-interactions and its 
interaction with the inflaton field, and find that DM can be over produced 
in a significant region of the parameter space. We further demonstrate that 
large self-interaction of the DM can suppress its abundance during preheating
and to a certain extent helps to achieve the observed relic abundance via cannibalization.
}

\begin{document}
\begin{flushright}
	PI/UAN-2018-638FT
\end{flushright}
\maketitle

\section{Introduction}
The existence of non-luminous Dark Matter (DM) has been well established 
thanks to several evidences in cosmology and astrophysics at different 
scales~\cite{Bertone:2010zza}. While these signatures only require DM to posses 
gravitational interactions and to be non-relativistic sufficiently early,  
the possibility that it can interact weakly with the Standard Model (SM) 
particles have been widely considered in various well motivated extensions 
of the SM~\cite{Jungman:1995df,Bertone:2004pz,Bertone:2010zza}. 

However, very little is known about the particle properties of DM and its interactions 
with the SM particles. It is generally assumed that DM consists of Weakly Interacting Massive Particles (WIMPs), which, as the name suggests, weakly interact 
with SM particles~\cite{Arcadi:2017kky}. This assumption opens up the the possibility that DM 
particles were in thermal equilibrium with the SM in the early Universe; 
and thus, were produced via the standard freeze-out mechanism. Further, such 
an assumption provided several avenues to search for DM. However, the various direct~\cite{Akerib:2016vxi, Cui:2017nnn, Aprile:2018dbl} 
and indirect~\cite{Aguilar:2016kjl,Fermi-LAT:2016uux} DM detectors, as well as the collider experiments~\cite{Aaboud:2016obm, Sirunyan:2017hnk},
have not measured any of such interactions so far. The most constraining 
limits for WIMP DM come from the direct detection experiments, especially for DM masses 
in the GeV ballpark. 

Moreover, while the large scale structure and Lyman-$\alpha$ data~\cite{Baur:2015jsy, Irsic:2017ixq} 
constraint the mass of the DM particle to be greater than $\mathcal{O}(10)$~keV, 
the success of Big Bang Nucleosynthesis (BBN) prevents any non-SM relativistic 
degree of freedom to be in thermal equilibrium with the SM particles at the 
onset of BBN ($\simeq 1$~MeV)~\cite{Cyburt:2015mya}. This clearly disfavors a DM candidate 
with mass $\lesssim 1$~MeV which thermalizes with the SM particles.% 
\footnote{However, in certain scenarios it is possible to accommodate such a 
light thermal DM invoking additional interactions~\cite{Berlin:2017ftj,Dutra:2018gmv,Berlin:2018ztp}.} 
 
While, in no way the current searches exclude the WIMP scenarios in its entirety, 
it may be worth exploring different kind of DM candidates which feebly 
interact with the SM particles. Generically such DM candidates do not 
thermalize with the SM particles in the early Universe, due to the feeble 
interaction strength. Several such scenarios have been proposed~\cite{McDonald:2001vt,Choi:2005vq,Kusenko:2006rh,Petraki:2007gq,Hall:2009bx}, both 
with a simple DM candidate, or even with a more complex Dark Sector (DS). Since in all these scenarios, DM is initially not in thermal equilibrium 
with the SM particles, the production mechanisms usually involve assumptions 
on the initial abundance of the DM or DS particles in the very early Universe. 
For example, in the simplest 
Feebly Interacting Massive Particle (FIMP) scenario, the initial abundance of 
the FIMP DM is usually assumed to be zero (see Ref.~\cite{Bernal:2017kxu} for a recent review).
%, the same holds for scenarios using re-annihilation, while in the scenarios using dark freeze-out make assumptions regarding the relative entropy of the DS with respect to the SM (see Ref.~\cite{Bernal:2017kxu} for a recent review). 
Therefore, an estimation of the initial abundance of DM or DS particles 
can be important in establishing the viability of a particular DM scenario.

In the very early Universe, the paradigm of cosmological inflation~\cite{Starobinsky:1980te,Linde:1981mu,PhysRevLett.48.1220} 
has been well-established, single field slow-roll inflation scenario have been 
very successful in the light of present data~\cite{Akrami:2018odb}. 
In this framework, post-inflationary particle production takes place during the 
(p)reheating epoch~\cite{PhysRevD.42.2491,  PhysRevLett.73.3195, PhysRevD.51.5438, PhysRevD.56.3258}. While during reheating particles are produced due to  
perturbative decay of the inflaton, preheating is a fast and efficient 
process, when particles are produced due to parametric resonance. This 
takes place even before the reheating epoch. Thanks to the Bose enhancement, 
this process is especially efficient in producing bosons which couple 
to the inflation. 

In the present article a simple extension of the SM, accommodating 
an inflaton and a scalar DM (stabilized using an effective $\mathbb{Z}_2$ symmetry), has been considered. 
Further, DM is assumed to posses negligible interaction with SM particles. 
Within this framework, we have studied the production of DM 
after inflation, during the epoch of preheating and reheating. The role of 
various interaction terms between the inflaton and the DM, as well as the
DM self-interactions,  have been 
analyzed in details. 

This article is organized as follows. In section~\ref{model} the model is presented, making especial emphasis on requirements for driving the inflation. 
In the following section~\ref{constraints} existing constraints on the relevant
model parameters various cosmological and astrophysical considerations have been 
sketched. Subsequently, in section~\ref{results} we discuss the production 
of the scalar DM during (p)reheating and estimate the relic abundance of the DM. 
Finally, in section~\ref{conclusion} we summarize our findings.

\section{The Model}\label{model}

The production of a scalar DM and SM particles during (p)reheating epoch 
generally depends on the inflaton potential at the end of inflation, and 
also on the interaction terms between the inflaton and the respective 
sectors. In this section, we sketch the model of our interest, elaborating 
on the inflationary scenario and the relevant interaction terms. 

On top of the SM field content, our model contains two real scalar fields, the 
inflaton $\phi$ and the DM $\chi$.\footnote{Note that a stable inflaton could also play the role of DM~\cite{Liddle:2006qz,Cardenas:2007xh,Panotopoulos:2007ri,Liddle:2008bm,Bose:2009kc,Lerner:2009xg,DeSantiago:2011qb,Khoze:2013uia,Mukaida:2014kpa,Fairbairn:2014zta,Bastero-Gil:2015lga,Kahlhoefer:2015jma,Tenkanen:2016twd,Daido:2017wwb,Choubey:2017hsq,Daido:2017tbr,Hooper:2018buz,Borah:2018rca,Manso:2018cba,Rosa:2018iff,Almeida:2018oid}.}
In order to ensure the stability of the DM, a $\mathbb{Z}_2$ symmetry 
is imposed under which only the DM is odd. The scalar potential is, then, 
given by 
\begin{eqnarray}\label{potential0}
	V &=& \frac{m_{\phi}^2}{2}\phi^2+\frac{m_\chi^2}{2}\chi^2-\frac{\mu_H^2}{2}|H|^2+\frac{\sigma_3}{3} \phi^3+\frac{\lambda_{ \phi}}{4}\phi^4  + \frac{\lambda_{\chi}}{4}\chi^4 + \frac{\lambda_H}{4} |H|^4\nonumber\\
	&&+ \frac{\sigma_{\phi \chi}}{2} \phi \chi ^2 + \frac{\sigma_{\phi H}}{2} \phi |H|^2 + \frac{\lambda_{\phi \chi}}{2} \phi^2 \chi ^2+ \frac{\lambda_{\phi H}}{2} \phi^2 |H|^2 + \frac{\lambda_{\chi H}}{2} \chi^2 |H|^2+V_\text{gravity}\,,
\end{eqnarray}
where $H$ is the SM Higgs field. Note that $\mu_H$ and $\lambda_H$, at the 
electroweak scale, are constrained in order to reproduce the observed Higgs 
mass and vacuum expectation value ({\it vev}).

In the present context, we will consider 
large field inflationary models. Two such scenarios have been widely studied, 
one with the quartic potential~\cite{Linde:1983gd} and another with quadratic 
potential~\cite{Mukhanov:1981xt} for the inflaton field $\phi$, see also 
Refs.~\cite{Linde:2007fr,Martin:2013tda} for reviews. In their simplest incarnations, both of 
these scenarios have been in tension with the present constraints from 
the CMB data~\cite{Akrami:2018odb}, the quartic one being worse. However, it has been 
pointed out that introduction of a non-minimal coupling with 
gravity can render both these scenarios viable~\cite{Fakir:1990eg,Linde:2011nh,Kallosh:2013tua}. While we will consider 
quadratic inflation in the presence of a non-minimal gravitational coupling, 
we will also comment on the consequences of considering a quartic potential, 
especially in the context of DM production during (p)reheating.
The non-minimal couplings to gravity are expressed as 
\begin{equation}
	V_\text{gravity}=\frac12\left(\xi_\phi\phi^2+\xi_\chi\chi^2+2\xi_H|H|^2\right)\mathcal{R},
\end{equation}
where $\mathcal{R}$ is the Ricci scalar~\cite{Birrell:1982ix}. For simplicity, we will set $\xi_\chi = 0 = \xi_{H}$.

Before discussing more on the inflationary aspects, which include terms depending 
only on the inflaton $\phi$, a discussion on various 
other relevant parameters are in order. As we will see in section~\ref{results},
they play an important role during the (p)reheating.
The trilinear terms $\frac{\sigma_{\phi \chi}}{2} \phi \chi ^2 $, $\frac{\sigma_{\phi 
H}}{2} \phi |H|^2$ and $\frac{\sigma_3}{3} \phi^3$ can potentially give rise to a
non-zero {\it vev} for the inflaton field at the end of inflation.%
\footnote{These terms are allowed by symmetries and required for the draining of the  
excess energy stored in inflaton after preheating, as 2-to-2 scatterings of inflatons 
into DM or SM through terms $\frac{\lambda_{\phi \chi}}{2} \phi^2 \chi ^2$ and $
\frac{\lambda_{\phi H}}{2} \phi^2 |H|^2$ can not transfer the energy stored in 
inflaton completely.} This will generate an effective mass term, presumably large, 
for both the scalars coupled to the inflaton. In order to keep the (p)reheating 
dynamics simple, and the DM mass as a free parameter of the theory 
(independent of $\lambda_{\phi\chi}$), we discard this possibility. Further, 
we also ensure that the minima of the scalar potential is at zero, and 
is achieved when all the scalar fields assume zero {\it vev}.%
\footnote{Since we are interested in the the epoch after inflation, where all 
relevant scales are much larger compared to the electroweak scale, we ignore 
the zero-temperature electroweak {\it vev} of the Higgs field.} 
These lead to the conditions $\lambda_{\chi}>\frac{\sigma_{\phi \chi}^2}
{2\alpha^2m_{\phi}^2}$, $\lambda_{H}>\frac{\sigma_{\phi H}^2}{2\beta^2m_{\phi}^2}$ 
and $\lambda_{\phi}>\frac{2\sigma_3^2}{9\gamma^2 m_{\phi}^2}$ given arbitrary real 
constants $\alpha$, $\beta$ and $\gamma$, such that $\alpha^2+\beta^2+\gamma^2=1$. 
Also the quartic couplings $\lambda_{\phi \chi}$, $\lambda_{ \phi H}$ and 
$\lambda_{\chi H}$ need to be positive. Note that at high scales ($\sim 10^{-5}M_\text{Pl}$) 
the quartic coupling $\lambda_H$ becomes negative, due to the large 
quantum corrections mostly from the top quark Yukawa coupling.% 
\footnote{Since in our context both $\lambda_{\phi \chi}$ and $\lambda_{\phi H}$ 
remain small, as we will discuss latter, the quantum contribution from the 
scalar fields $\phi$ and $\chi$ do not improve the situation.} 
This has been well studied in the literature, see e.g. Refs.~\cite{Sher:1988mj,EliasMiro:2011aa,Degrassi:2012ry,Buttazzo:2013uya}. 
Its consequence during inflation and possible remedies are also known,
see e.g. Refs.~\cite{Espinosa:2007qp,Enqvist:2014bua,Shkerin:2015exa,Kearney:2015vba,Espinosa:2015qea}
(also Ref.~\cite{Espinosa:2007qp,Herranen:2014cua,Herranen:2015ima,Ema:2017loe} for $\xi_H \neq 0$). 
Since our main focus is on the post inflationary production of DM particle 
during (p)reheating, we will assume that 
the quartic coupling $\lambda_H$ remains positive during (p)reheating epoch. 
This can be simply achieved by introducing another scalar field, 
which couples to the Higgs boson~\cite{Lebedev:2012zw,EliasMiro:2012ay}.
We will assume that such a scalar, if exists, 
thermalizes with the SM particles, and if coupled to the inflaton, its 
production and dynamics during (p)reheating would be similar as that of the 
Higgs boson itself; but we will not explicitly include it in our discussion. For 
the rest of this article, we will simply assume that during (p)reheating 
$\lambda_H$ remains positive.

For the inflationary paradigm we consider the quadratic 
potential for the inflaton $\phi$. In this case, in Eq.~\eqref{potential0} the quadratic 
term for the inflaton 
field $\phi$ dominates over the quartic term, which is assumed to be 
small. Introducing a non-minimal coupling $\xi_{\phi} \simeq \mathcal{O}(10^{-3})$ 
with $m_\phi \simeq \mathcal{O}(10^{-6})~M_\text{Pl}$~\cite{Linde:2011nh} 
the quadratic potential can produce the scalar spectral index $n_s$, the tensor-to-scalar ratio 
$r$ and the amplitude of the scalar perturbation $A_s$, consistently with the current Planck data~\cite{Akrami:2018odb}. 
In particular, a benchmark with $\xi_\phi=2\times 10^{-3}$, $m_{\phi}=5.3\times10^{-6}~M_\text{Pl}$ 
(with $\phi_{\rm pivot} \simeq 14\,M_\text{Pl}$ and $\phi_{\rm end} 
\simeq M_\text{Pl}$), leads to the following values of the observable parameters: 
the scalar spectral index $n_s \simeq 0.963$, the amplitude of the scalar 
perturbation $A_s \simeq 2.22\times 10^{-9} $ and the tensor-to-scalar ratio $r \simeq 0.043$, while the 
inflation lasts for  $\sim 62~e$-foldings. It has also been shown in literatures \cite{Markkanen:2015xuw,Fairbairn:2018bsw} that a non-minimal coupling of $\chi$ to gravity can produce DM gravitationally, even if the DM has no other coupling with any other sector.

Note that, the success of inflation critically depends on the flatness 
of the inflaton potential, such that the slow-roll parameters remains 
small during the entire inflation. In case of the quadratic inflation, 
the contribution from the quartic term, therefore, must remain insignificant. 
Note that, simply forbidding $\lambda_{\phi}$ at the tree-level, would not 
suffice, since it can be generated due to quantum corrections. The relevant renormalization group equations (RGE) are given by~\cite{DeSimone:2008ei,Lerner:2011ge,Ema:2017ckf}, 
\begin{eqnarray}
	16\pi^2 \frac{d \lambda_{\phi}} {d\ln{\mu}} &=& 8 \lambda_{\phi H}^2 + 2 \lambda_{\phi \chi}^2+ 18 c_\phi^2\lambda_{\phi} ^2\,,\label{eq:rge1}\\
	16\pi^2 \frac{d \lambda_{\phi H}}{d\ln{\mu}} &=&\lambda_{ \phi H}\left[8c_\phi\lambda_{ \phi H} + 12 \lambda_H+6c_\phi^2\lambda_\phi-\frac32(3g_L^2+g_Y^2)+6y_t^2\right],\label{eq:rge2}\\
	16\pi^2 \frac{d \xi_\phi}{d\ln{\mu}} &=&6\left(\xi_\phi-\frac16\right)c_\phi\lambda_\phi,\label{eq:rge3}
\end{eqnarray}
where
\begin{equation}
	c_\phi=\frac{1+\xi_\phi \phi_i^2/M_\text{Pl}^2}{1+(6\xi_\phi-1)\,\xi_\phi \phi_i^2/M_\text{Pl}^2},
\end{equation}
with $g_L$ and $g_Y$ corresponding to the $SU(2)$ and $U(1)$ gauge couplings and $y_t$ being the Yukawa coupling of the top quark. As shown in the equations above, the radiative contribution to $\lambda_{\phi}$ 
involves both $\lambda_{\phi \chi}$ and $\lambda_{\phi H}$. Thus, for requiring 
$\lambda_{\phi} \lesssim 10^{-14}$ (to ensure the smallness of this term during inflation) without invoking any tuning, we assume that $\lambda_{\phi \chi}$,
$\lambda_{\phi H} \lesssim \mathcal{O}(10^{-7})$ during inflation and right after 
the inflation at the onset of (p)reheating epoch. Similar argument prevents 
$\sigma_3$ from being large. Further, $\sigma_{\phi \chi}$ and $\sigma_{\phi H}$ 
can also contribute to the RGE of $m_{\phi}$ and $\sigma_3$, and we ensure 
that these parameters remain small during (p)reheating after inflation in our 
discussion in section~\ref{results}.

Finally, we comment on the possibility of considering the quartic inflation 
where the quadratic term $\frac{m_{\phi}^2}{2} \phi^2$ is negligible in Eq.~\eqref{potential0}
compared to $\frac{\lambda_{\phi}}{4} \phi^4$ during inflation~\cite{Fakir:1990eg,Makino:1991sg,Komatsu:1997hv,Komatsu:1999mt,Tenkanen:2017jih}. 
However, to ensure that $\phi$ remains massive after inflation (without the necessity 
of a phase transition), and to ensure that the minima of the scalar potential 
remains at zero, when all the scalar fields are set to zero, a mass term for 
$m_{\phi}\ll 10^{-6} M_{\rm Pl}$ would be desired. It has been shown that, 
in this case, the current CMB data can be matched with $N\sim 60~e$-folds and $\xi_\phi\sim 790N\sqrt{\lambda_\phi}$.
%$\xi_{\phi} \simeq 10^{-3}-1$ and $\lambda_{\phi} \simeq 10^{-13}-10^{-10}$~\cite{Ballesteros:2016xej}. 
Note that the nature of the inflaton potential, whether quartic or quadratic, 
affects the post inflation dynamics of the inflaton, and hence the preheating 
process. Furthermore, a heavy inflaton can possibly behave as matter after the 
rapid preheating process, before it decays injecting significant entropy, 
and diluting the abundance of any light particle produced during 
preheating. In section~\ref{results}, a quadratic potential will be assumed 
for $\phi$. However, we will also comment on the impact of the choice of 
quartic inflation in the same context.

 \section{Constraints}\label{constraints}
In this section, we briefly discuss the relevant constraints, especially focusing on a light DM candidate.

$\bullet$ While only active neutrinos lead to $N_\text{eff}=3.046$~\cite{Mangano:2005cc}, the presence 
of any additional relativistic species enhances $N_\text{eff}$. Primordial nucleosynthesis yields, independently of measurement of the baryon density from CMB observations, $\Delta N_\text{eff}<1$ at 95\%~CL~\cite{Mangano:2011ar}. Using CMB measurements along with the BBN yields data can improve the constraint to $\Delta N_\text{eff}<0.38$ at 95\% CL~\cite{Cyburt:2015mya} (see also Refs.~\cite{Cooke:2017cwo, Tanabashi:2018oca}). This is a constraint we keep in mind when having a DM of the mass range keV-MeV (i.e. relativistic during BBN). To satisfy it, the ratio of dark sector energy density to energy density of relativistic SM particles is required to be $\lesssim 0.051$ at the onset of BBN \cite{Cyburt:2015mya}.
 
$\bullet$ We consider DM particles with mass range of minimum $m_\chi\gtrsim \mathcal{O}(10)$~keV as constrained by Lyman-$\alpha$ observations~\cite{Baur:2015jsy, Irsic:2017ixq}.
 
$\bullet$ The presence of a light scalar field, the DM, 
during inflation can leave imprints in the primordial power spectrum. In the 
context of curvaton scenarios~\cite{PhysRevD.42.313, Linde:1996gt,Enqvist:2001zp,Lyth:2001nq,Moroi:2001ct} these possibilities have been considered. 
The generic signatures include significant non-Gaussianity and isocurvature 
modes in the primordial power spectrum.%
\footnote{Note that, in the context of multi-field inflation scenarios, where more 
than one fields contribute to the exponential expansion with comparable 
energy densities, isocurvature perturbation can also arise~\cite{POLARSKI1992623,Langlois:1999dw,Gordon:2000hv,Lalak:2007vi}.} 
Recent results from Planck~\cite{Akrami:2018odb} find no evidence for 
isocurvature perturbation, imposing stringent constraints. The 
success of single field slow-roll inflation paradigm in the light of present 
CMB data implies that such an additional light field, if present during inflation, 
can only contribute subdominantly to the energy density of the Universe during 
inflation. For a light field during inflation, $m^\text{eff}_{\chi} \ll H_{\rm inf}$, 
where $m^\text{eff}_{\chi}$ and $H_{\rm inf}$ denote the effective mass of the light 
scalar and Hubble parameter during inflation, respectively. The light field will 
acquire quantum fluctuations proportional to $H_{\rm inf}$~\cite{Linde:1982uu, Starobinsky:1982ee, Vilenkin:1982wt,Starobinsky:1994bd} with its root mean squared value given by 
$\frac{\mathcal{O}(0.1)}{\lambda_\chi^{1/4}}\,H_{\rm inf}$~\cite{Starobinsky:1994bd}. 
The condensate, thus formed, remains frozen until the end of inflation while the 
Hubble parameter becomes small compared to the (possibly thermal) mass of the 
light field. At this point the field starts to oscillate and behaves like 
matter before finally decaying into the $\chi$ particles
(in the quartic region~\cite{Kainulainen:2016vzv}). Note that, since we assume very small coupling 
with the SM fields (i.e. negligible portal coupling with the Higgs boson), $\chi$ 
particles never thermalize with the SM sector. Similar ideas have been studied 
in Refs.~\cite{Enqvist:2014zqa,Nurmi:2015ema,Kainulainen:2016vzv}.
In Ref.~\cite{Kainulainen:2016vzv} it has been shown that, such a scenario would be 
quite constrained from non-observation of isocurvature perturbation, 
in particular, for large field inflation models (with large $H_{\rm inf} \simeq 10^{13}-10^{14}$ GeV). In order to evade this issue, as we will elaborate in 
the subsequent section, we consider sizable coupling of the $\chi$ field with 
the inflaton $\phi$. This enhances the effective mass (i.e. the second derivative of 
the relevant potential) of the $\chi$ field during inflation and ensures that the 
field does not receive large fluctuations during the same period.  

$\bullet$ We demand a successful inflation, able to reproduce the observed scalar spectral index $n_s$, the tensor-to-scalar ratio $r$ and the amplitude of the scalar perturbation $A_s$.

$\bullet$ We finally demand to reproduce the observed DM relic abundance.

\section{Dark Matter Production}\label{results}

\begin{figure}[t!]
	\centering
	\includegraphics[width=0.475\linewidth]{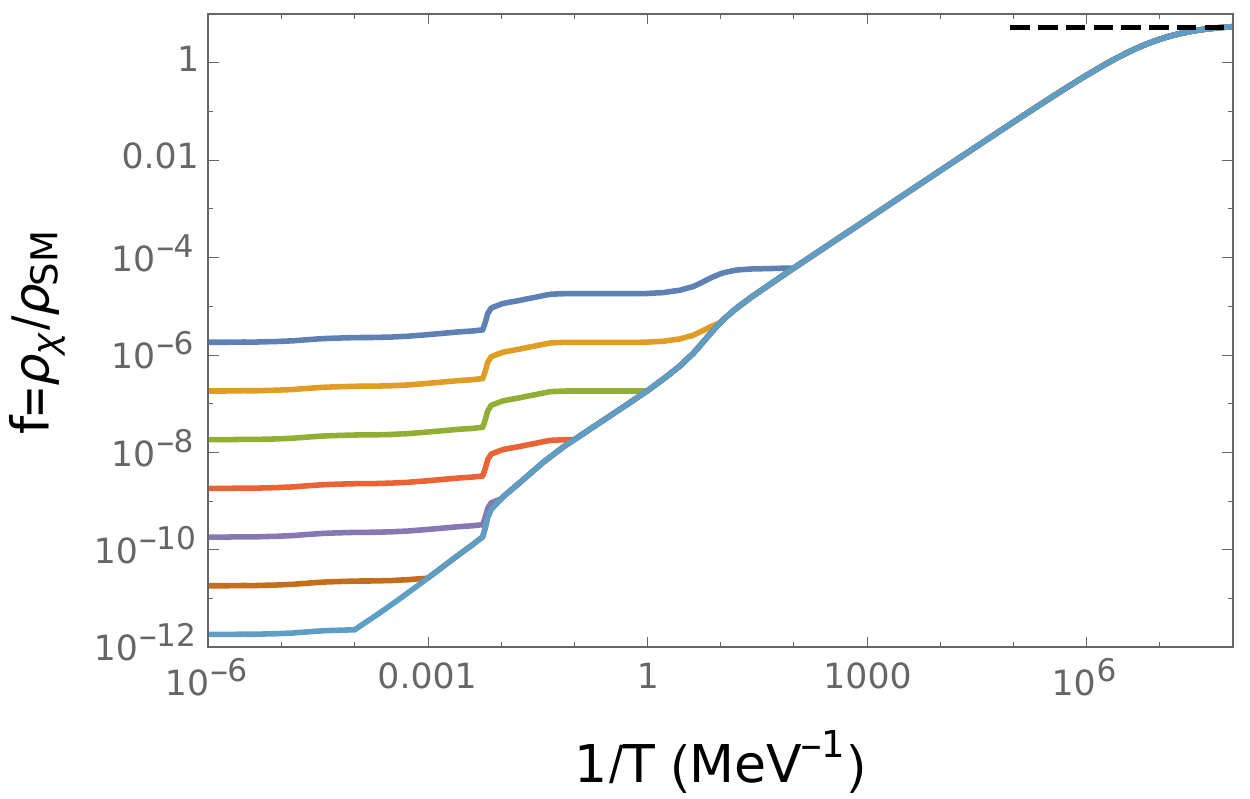}
	\caption{Energy fraction $f(T)\equiv\frac{\rho_\chi(T')}{\rho_\text{SM}(T)}$ required to 
	get the observed DM abundance as a function of $1/T$ (in MeV$^{-1}$), for $m_\chi=0.01,~0.05,~0.3,~2,~10,~50$ and $300$~MeV (from top to bottom). The dashed line at 
	the top-right corner corresponds to the observed $f(T_0)\sim 5.3$, as measured by Planck.
	}
	\label{bottomtop}
\end{figure}

The ratio of the energy density of the DM to that of the SM at a temperature $T$, 
$f(T)\equiv\frac{\rho_\chi(T')}{\rho_\text{SM}(T)}$, measured at the present day CMB 
temperature $T_0$ is $f(T_0)\sim 5.3$~\cite{Aghanim:2018eyx} (and a corresponding DM temperature $T'$).
In the early Universe, however, this ratio is expected to be much smaller since 
SM radiation cools down faster than the DM component.
Assuming the two sectors were always completely disconnected, with corresponding entropies separately conserved, the evolution of $f$ is shows in Fig.~\ref{bottomtop}, for different DM masses: $m_\chi=0.01,~0.05,~0.3,~2,~10,~50$ and $300$~MeV  (from top to bottom).
The initial values (at high temperatures) for all curves are tuned so that at low temperatures they reproduce the observed DM abundance, as measured by Planck, and showed with a dashed line.
When $T'\gtrsim m_\chi$, $f$ tends to be constant showing that both components scale in the same way, as radiation.
An analog behavior appears for $T<0.7$~eV, near the matter-radiation equality, when both the SM and the DM behave like matter.
The kinks at $T\sim 0.1$~MeV and $T\sim 100$~MeV correspond to the electron-positron annihilation and the QCD phase transition, respectively.

In the following we will focus on how (p)reheating could generate the DM to SM energy density ratio in order to reproduce the observed DM abundance, firstly assuming that there are no sizable self-interactions for the DM, and secondly taking them 
into account. 

%%%%%%%%%%%%%%%%%%%%%%%%%%%%%%%%%%%%%%%%%%%%%%%%%%%%%%%%%%%%%%%%%%%
\subsection{Dark Matter Production during (P)reheating}
In this subsection we consider the DM production after inflation during the 
(p)reheating epoch. The process of reheating involves perturbative decay of 
the inflaton and the production of a particular species during this epoch 
can be estimated by the branching fractions of the inflaton into it. However,
during preheating, which precedes the reheating epoch, bosonic particles coupled 
to the inflaton field are produced via parametric or tachyonic resonance, and 
it is a rapid non-perturbative process. The initial stage of preheating can 
be described in terms of a set of Mathieu equations in the Fourier space 
and can be studied analytically~\cite{PhysRevLett.73.3195}. The Fourier modes, 
which are present in the well-known instability bands, grow exponentially at 
the onset of preheating and are interpreted as exponential particle production%
~\cite{PhysRevLett.73.3195}. However, after the initial stage of preheating, 
non-linear effects including back-reaction and re-scatterings 
become relevant~\cite{PhysRevD.56.3258}. Thus, during this phase, accurate analytical estimations are 
hard to obtain. Therefore, numerical codes have been developed and used to
estimate the particle production during this epoch. We have used the publicly 
available code {\tt LATTICEEASY}~\cite{Felder:2000hq} in order to the 
study the DM production during preheating.
The inflationary production of scalar DM with intermediate masses has been studied by various authors~\cite{Dev:2013yza,Nurmi:2015ema,Kainulainen:2016vzv,Bertolami:2016ywc,Heurtier:2017nwl,Cosme:2017cxk,Cosme:2018nly,Alonso-Alvarez:2018tus}.

\begin{figure}[t!]
	\centering
	\includegraphics[width=0.475\linewidth]{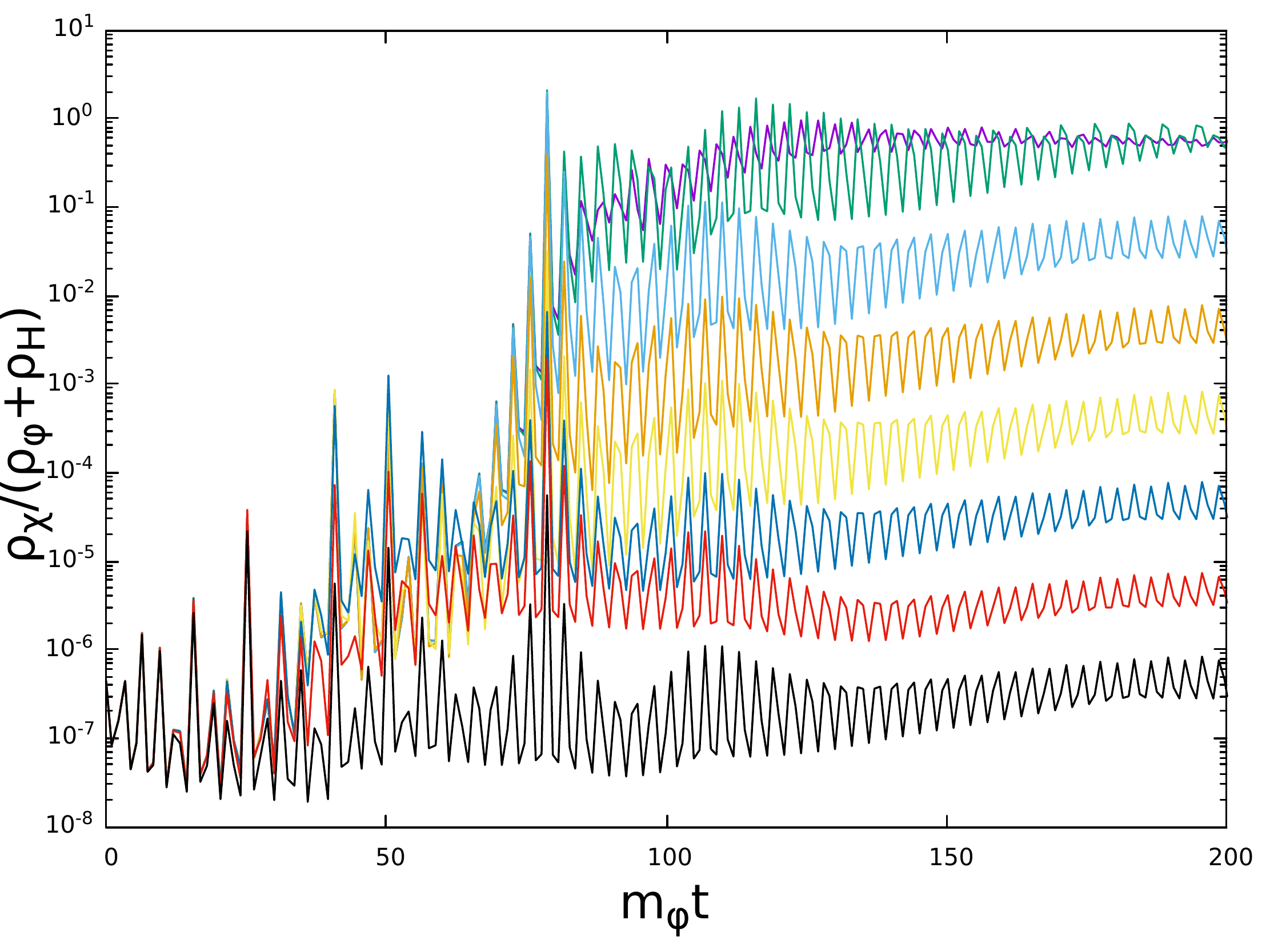}
	\caption{Evolution of the energy fraction $\rho_\chi/(\rho_\phi+\rho_H)$ as a function of time for $m_{\phi}=10^{-6}M_\text{Pl}$, $\lambda_{\phi}=10^{-14}$, $\lambda_{\phi\chi}=10^{-7}$, $\lambda_{\phi H}=10^{-7}$, $\lambda_{H}=10^{-7}$ and $\sigma_{\phi H}=10^{-10}M_\text{Pl}$. $\lambda_{\chi}=10^i$, for $i=-7$ (uppermost line) to $i=0$ (lowermost line), increasing $i$ in one unit each time. Higher values of $\lambda_\chi$ give rise to more suppression of DM energy density.
	}
	\label{ratioplot}
\end{figure}

In Fig.~\ref{ratioplot}, we plot the evolution during preheating of the ratio of energy density of $\chi$ ($\rho_{\chi}$)  
to the combined energy density of $\phi$ ($\rho_{\phi}$) and Higgs ($\rho_{H}$) together, i.e. $\dfrac{\rho_{\chi}}{\rho_{\phi}+\rho_{H}}$.  We vary $\lambda_{\chi}$ while keeping the other parameters fixed 
as follows: In order to ensure a successful inflationary phase, $m_\phi=10^{-6}~M_{\rm Pl}$ and $\lambda_\phi=10^{-14}$. Further, $\lambda_{\phi\chi}=\lambda_{\phi H}=10^{-7}$ such that  the flatness of the inflationary potential is not ruined, as discussed in section~\ref{model}. A small $\lambda_H=10^{-7}$, as we will elaborate in the subsequent 
discussion, avoids the suppression of the energy flow into Higgs during preheating. Finally, we have assumed $\sigma_{\phi H}=10^{-10}M_\text{Pl}$ respecting 
the relation $\lambda_{H}>\frac{\sigma_{\phi H}^2}{2\beta^2m_{\phi}^2}$, again 
as discussed in section~\ref{model}, while $\sigma_{\chi H} = 0$ in order to avoid 
thermalization of $\chi$ and Higgs. In the figure, $m_\phi t=0$ denotes the onset of preheating at the end of 
inflation where we set $\phi_{ini} \simeq 0.2\,M_{\rm Pl}$ at which \texttt{LATTICEEASY} 
starts to evolve the preheating dynamics. While for different values of $\lambda_{\chi}$ 
while the ratio changes significantly, as shown in Fig.~\ref{ratioplot}, the generic 
features of the curves resemble each other, since these are dependent on the dynamics of 
preheating in general. As emphasized already, production of $\chi$ and $H$ quanta 
during the initial stage of preheating is governed by the relevant Mathieu equation 
and the energy transfer from the oscillating inflaton condensate to the $\chi$ and $H$ sector is significant. The energy transfer is most efficient when the inflaton is 
near the minimum of the potential, i.e. $\phi(t) \simeq 0$, since during this phase 
the adiabatic conditions are strongly violated (due to large $\dot{\phi}$) and also 
the contribution of $\phi(t)(\simeq 0)$ to the effective mass of $\chi$ and $H$ 
becomes small. While the production of $\chi$ is dominated by the parametric resonance
due to the absence of the trilinear term $\sigma_{\phi \chi}$, the production 
of $H$ can receive contributions due to both parametric and tachyonic resonances 
(for small $k$ modes in particular, in our context)~\cite{Dufaux:2006ee}. The 
oscillatory features in the energy densities (and in their ratio, as shown in 
Fig.~\ref{ratioplot}) during the initial stage of preheating can be attributed 
to the characteristic solution of Mathieu equation. However, as the $\chi$ and 
$H$ fields are populated, as discussed in the previous paragraph, the 
back-reaction and re-scattering become significant, gradually suppressing the 
energy transfer to the $\chi$ and $H$ fields, and the ratio $\dfrac{\rho_{\chi}}{\rho_{\phi}+\rho_{H}}$ eventually stabilizes.% 
\footnote{In particular, once in the initial stage the resonant modes $k_*$ of $\chi$  
($H$) get excited, annihilation of those quanta generate inflaton quanta, re-scattering 
of $\chi$  ($H$) quanta against inflaton zero mode produces $\chi$ ($H$) and inflaton 
quanta with the momentum $k_*/2$. In the next stage, violent phase of non-linear 
dynamics, the $k_*/2$ modes grow in amplitude and shift towards $k_*$. In the third 
stage, the distribution smooths out and spreads towards higher momentas. For an elaborate 
discussion, see Ref.~\cite{Podolsky:2005bw}.}
For example, as shown in the figure, with $\lambda_\chi = 10^{-7}$ (the highest line) 
the increment continues until $m_\phi t\sim80$, at which point $\rho_{\chi}$ becomes 
comparable to that of the inflaton and does not grow any further. 
Note that, with $\lambda_{\phi H} \simeq \lambda_{\phi \chi}$ (both set to 
$\mathcal{O}(10^{-7})$), %respecting the constraints, as argued in the previous section),
as the  self-coupling of $\chi$ ($\lambda_{\chi}$) is enhanced compared to 
that of $H$, the energy flow in the $\chi$ sector decreases substantially. In 
particular, as shown in the figure, for higher quartic couplings: $\lambda_{\chi}=10^0$ 
(lowermost line), $10^{-1}$, ... $10^{-6}$ and $10^{-7}$ (uppermost line) the ratio 
$\dfrac{\rho_{\chi}}{\rho_{\phi}+\rho_{H}}$ varies from $\mathcal{O}(10^{-6})$ to $\mathcal{O}
(10^{-1})$, respectively. A similar suppression due to quartic self-interaction has 
also been observed in Refs.~\cite{Prokopec:1996rr, Hyde:2015gwa}. This feature 
can be attributed to the fact that the quartic coupling, after initial exponential 
production of DM, contributes to the effective mass of $\chi$ as $\sqrt{\frac{\lambda_{\chi}}{2}\langle \chi^2\rangle}$. This large contribution 
makes subsequent production of $\chi$ energetically expensive, effectively blocking 
the same~\cite{Dufaux:2006ee}.
%It is worth mentioning here that, it has been pointed out in Ref.~\cite{Hardy:2017wkr} that (without considering the trilinear and the quartic terms for the relevant field) the production of a boson during preheating can be suppressed by reducing its interaction strength with the inflaton $\lambda_\chi \lesssim 4 \times 10^{-8}$ which leads to an inefficient and short period of preheating, due to the smallness of the relevant parameter $\lambda_\chi \phi_{ini}^2/m_{\phi}^2 \lesssim 1$.%
It is worth mentioning here that the production of a boson during preheating can be suppressed by reducing its interaction strength with the inflaton $\lambda_{\phi\chi}\lesssim 4 \times 10^{-8}$, which leads to an inefficient and short period of preheating, due to the smallness of the relevant parameter $\lambda_{\phi\chi}\,\phi_{ini}^2/m_{\phi}^2 \lesssim 1$~\cite{Hardy:2017wkr}.%
\footnote{However, this does not seem to lead 
to adequate suppression for different DM masses we have considered. Also, 
if the interaction is significantly reduced, the effective mass of the boson may fall below $H_{\rm inf}$. We will not consider this possibility.}  
Finally, although not explicitly shown in Fig.~\ref{ratioplot} at the end of 
preheating, the remaining energy density in the inflaton condensate continues 
to oscillate before it eventually decays, in our context, mostly producing 
the SM Higgs, thanks to the trilinear $\sigma_{\phi H}$ term.% 
\footnote{Note that during the oscillation of the inflaton, an effective 
trilinear coupling proportional to $\lambda_{\phi \chi} \phi(t)$ can be generated. 
Thus, setting $\sigma_{\phi \chi}=0$ does not completely prohibit the perturbative 
decay of the inflaton into $\chi$. However with time, as the amplitude of inflaton oscillation $\phi(t)$ decreases below $\mathcal{O}(10^{-3}M_{\rm Pl})$, 
$\sigma_{\phi H}\simeq \mathcal{O}(10^{-10} M_{\rm Pl})$ becomes dominant. For 
simplicity we have ignored this contribution in our estimation.}

While we are interested in the initial value of $f = \rho_\chi/\rho_\text{SM}$ at the 
end of (p)reheating, the decay mentioned above ensures that $\dfrac{\rho_{\chi}}
{\rho_{\phi}+\rho_{H}}$, as plotted in the figure above, gives a good estimate of 
$f$ at the end of (p)reheating assuming that the inflaton does not go through 
a non-relativistic phase before it decays. However, as shown in Fig.~\ref{bottomtop}, 
only a small initial DM energy density 
(compared to that of the SM) at high temperature suffices to produce the right 
DM relic abundance. Even with a rather large $\lambda_{\chi}$, and assuming that 
the perturbative decay of the inflaton $\phi$ only contributes to $\rho_\text{SM}$, 
we find it difficult to achieve such a small ratio, and produce substantially 
large initial abundance of $\chi$ at the end of preheating, especially for $m_{\chi} \gtrsim 0.02$ MeV. However, as will be discussed subsequently in this section, 
the mass range $m_{\chi} \lesssim 0.02$~MeV, giving rise to the right initial 
abundance with large $\lambda_{\chi}$, as discussed above, have been disfavored 
due to rather large self-interaction from the Bullet cluster bounds. 
This makes it apparent that the scenario, as described so far, requires 
mechanisms to suppress the initial abundance of $\chi$ in order to be viable.

It is well-known that the energy density associated to the oscillations of the 
inflaton, in the context of quadratic potential, behaves as matter. However, 
it has been shown that, although substantial energy density remains in the inflaton 
(condensate and the excitation combined) at the end of preheating, 
in presence of the trilinear $\sigma$ terms the equation of state right after 
preheating is rather close to that of the radiation $w \simeq \frac{1}{4}$~% 
\cite{Dufaux:2006ee}. This is in contrast to the case of quartic inflation, where 
the equation of state, under similar circumstances, resembles that of radiation.% 
\footnote{In Ref.~\cite{Dufaux:2006ee}, the equation of state during the preheating 
period has been studied in detail. It has been shown that for a preheating scenario (with 
quadratic inflaton potential) where there is no trilinear coupling of inflaton to other 
species, the equation of state at first increases to $1/4$, then it decreases slowly 
towards $0$. On the other hand, in a scenario with trilinear term, the equation of state 
increases to a value above $1/4$ and does not decrease thereafter. This phenomenon has 
been explained by studying the fraction of number density stored in the relativistic 
modes of the fields. It has been observed that, for the first scenario, it decreases 
after initial increment for the inflaton field. Whereas for the second scenario, this 
fraction remains stable for the entire simulation time scale. Therefore, for simplicity, 
we can assume the inflaton to be a relativistic species after preheating for our case 
with trilinear coupling.}

\begin{figure}[t!]
	\centering
	\includegraphics[width=0.475\linewidth,height=0.340\linewidth]{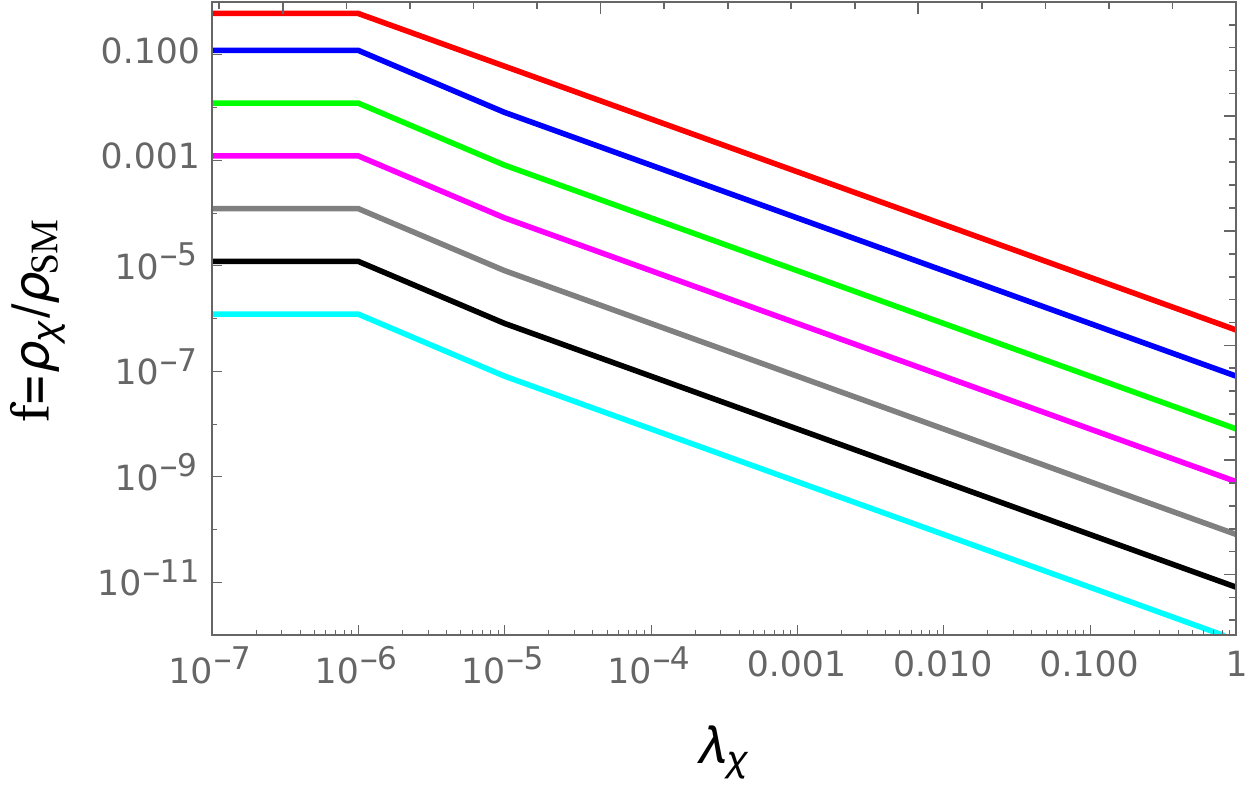}
	\caption{Energy fraction $f\equiv\rho_\chi/\rho_\text{SM}$ at the end of reheating as a function of $\lambda_{\chi}$, for the same benchmark point of Fig.~\ref{ratioplot}. The lines correspond to $a_d/a_t=10^0$ (uppermost) to $10^6$ (lowermost), increasing by an order of magnitude each time.
	}
	\label{ratioplot2}
\end{figure}

However, if inflaton becomes non-relativistic before it decays, its energy density after 
the relativistic to non-relativistic transition, scales as $a^{-3}$, while 
for the relativistic species (SM and DM particles) the energy density scales 
as $a^{-4}$, where $a$ is the scale factor.
This can lead to an enhancement in the relative energy fraction stored 
in the inflaton field at the end of preheating, and the subsequent decay process 
into SM can suppress the DM energy density relative to SM.
The dynamics of the Universe during the phase between the end of preheating (defined as the 
initial exponential particle production) and the decay of inflaton (governed by the decay 
width $\Gamma_\phi$ of inflaton, which is generally much larger than the preheating time 
scale) is complicated and contains processes like turbulence and thermalization~%
\cite{Aarts:2000mg, Allahverdi:2000ss, Felder:2000hr, Salle:2000hd, Allahverdi:2002pu, Micha:2003ws, Micha:2004bv, Desroche:2005yt, Mukaida:2015ria}. The enhancement of the relative energy density of inflaton from the 
end of preheating to its decay depends on details of this stage. % and is parametrized by a factor $a_d/a_t$, where $a_d$ and $a_t$ correspond to the scale factors when the inflaton decays and when it becomes non-relativistic. 
%For simplicity we parameterize our ignorance about this enhancement as a factor $a_d/a_t$, where $a_d$ and $a_t$ correspond to the scale factors when the inflaton decays and when it becomes non-relativistic 
%As we assume the other preheated sectors to be relativistic during the decay of inflaton, the factor can not be $a_d/a_t<1$.
Although the parameter $a_d/a_t$, written as a ratio of scale factors during the decay of inflaton $a_d$ and during the transition of inflaton energy density from relativistic to non-relativistic $a_t$ (where we assume that both the inflaton decay and its transition from relativistic to non-relativistic phase are instantaneous), enhances the inflaton energy density relative to the other (relativistic) energy densities, the full details of the phase  between end of preheating and decay of inflaton may add additional information to get the correct enhancement factor. We use $a_d/a_t$ as an effective enhancement factor of inflaton energy density with respect to other energy densities. 
The factor $a_d/a_t$ is by construction $\geq1$, as even if the inflaton decays when it is relativistic, i.e. $a_d< a_t$, the inflaton energy density does not get enhanced.
%If the inflaton decays after becoming non-relativistic, then there is a matter dominated phase, during which the energy density of the preheated species gets diluted.
In Fig.~\ref{ratioplot2}, we plot $f\equiv\frac{\rho_\chi}{\rho_\text{SM}}=\frac{\rho_\chi^{ini}}{\rho_\phi^{ini}\times\frac{a_d}{a_t}+\rho_H^{ini}}$ (where $\rho_i^{ini}$ is the initial energy density of species $i$ just after preheating) as a function of $\lambda_{\chi}$ at the end of preheating for different values of $a_d/a_t$.
For the top-most line of this plot, it is clear from Fig.~\ref{ratioplot} why the ratio does not change with $\lambda_{\chi}$, when $\lambda_{\chi}$ is not significantly larger than $\lambda_{\phi\chi}$, and after that the ratio falls off with increasing $\lambda_{\chi}$. The other lines in this plot correspond to the cases where the inflaton energy density gets a boost by a factor $a_d/a_t>1$. 

\begin{figure}[t!]
	\centering
	\includegraphics[width=0.475\linewidth]{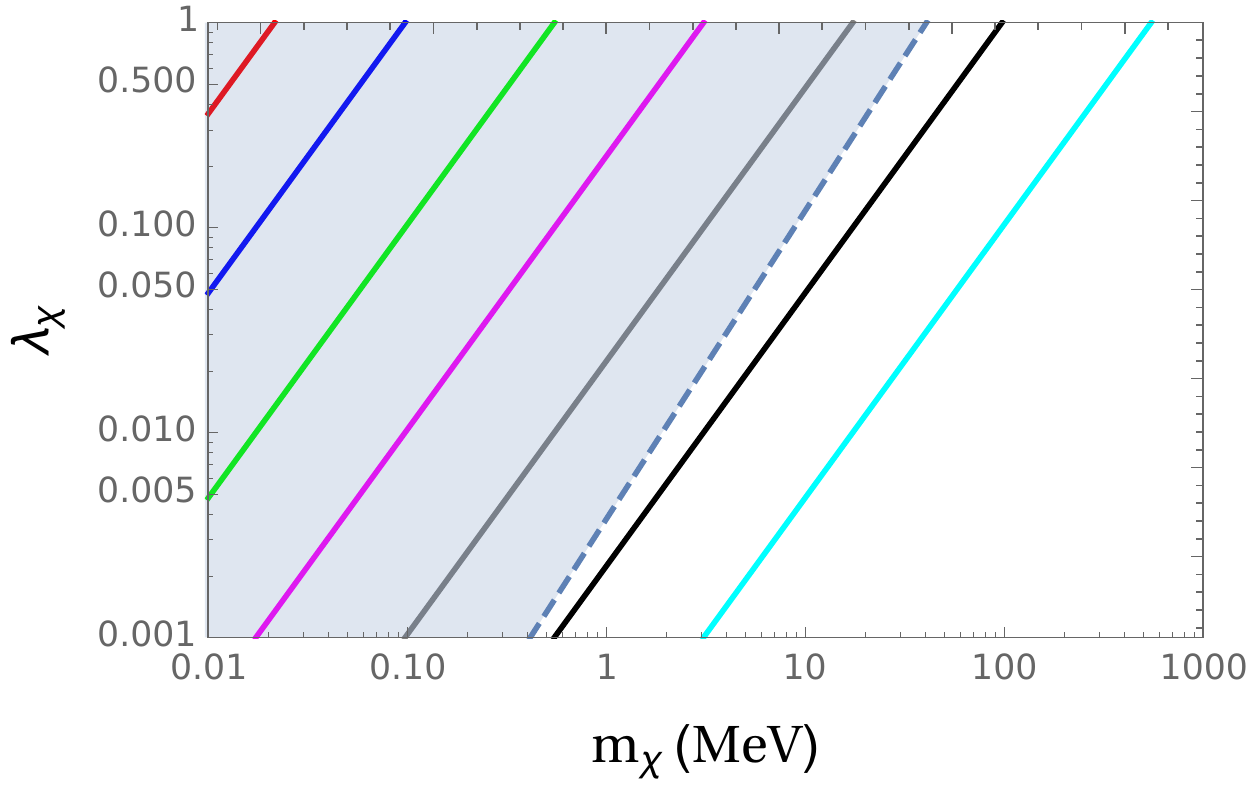}
	\caption{$\lambda_{\chi}$ vs $m_{\chi}$ (in MeV) plot for allowed relic abundance (without any depletion mechanism in DM sector). The lines from top-left corresponds to $a_d/a_t=10^0$ to $10^6$. The shaded region is excluded from Bullet cluster bounds.
	}
	\label{masl}
\end{figure}

Fig.~\ref{masl} combines the information of Figs.~\ref{bottomtop} and~\ref{ratioplot2}.
It depicts the values of the quartic coupling $\lambda_\chi$ required in order to reproduce the observed DM relic abundance as a function of the DM mass, for different ratios $a_d/a_t$: from $10^0$ (uppermost) until $10^6$ (lowermost), increasing by an order of magnitude each time.
The shaded gray region is in tension with the Bullet cluster constraint on DM self-interactions.
The figure shows that the case where $a_d/a_t=1$ is excluded by observations.
%If we consider $a_d/a_t=1$ and assume that there is no depletion mechanism of the $\chi$ sector, to get the right relic abundance of DM of some $m_\chi$, $f(m_\chi)$ (achieved from the high temperature region of Fig.~\ref{bottomtop}) we have to choose a $\lambda_{\chi}$ such that $f(m_\chi)=f(\lambda_{\chi})$ (where $f(\lambda_{\chi})$ is  the topmost line in Fig.~\ref{ratioplot2}).
%Fig.~\ref{masl} shows that obtaining correct relic abundance of DM from only suppression of preheating production by $\lambda_{\chi}$ is impossible (the line at the top-left corner which is basically the solution of the equation $f(m_\chi)=f(\lambda_{\chi})$, is totally ruled out by Bullet cluster constraints).
However, having $a_d/a_t\gg 1$ gives additional suppression of the DM to SM ratio if the inflaton fully decays into SM only.
An additional suppression in the ratio $f$ by this factor can reproduce the right DM abundance of a specific $m_\chi$ with a smaller value of $\lambda_{\chi}$.
Thus, as shown in Fig.~\ref{masl}, an appropriate $a_d/a_t$ may lead to the right DM abundance, evading Bullet cluster bounds.\footnote{Quartic inflation does not give a better result in terms of getting right DM relic abundance. This is because, the energy density of a quartic potential evolves as radiation while oscillating around its minima. For the same trilinear coupling strength, the inflaton may decay while still relativistic (because inflation may be rather light in this case), thereby possibly decreasing the duration of a non-relativistic phase.}
Let us emphasize that in this figure we are assuming that there are no interactions within the dark sector able to modify the DM number density.
However, when $\lambda_\chi$ becomes sizable, that may not be a suitable assumption.
In fact, in this case DM cannibalization will inevitably deplete the DM sector. This is especially efficient in the presence of large self-interaction parameter $\lambda_{\chi}$ which, incidentally,  also helps to suppress the DM production  during preheating. We will discuss this possibility in detail in the next subsection.

\subsection{Dark Matter Cannibalization}

In this scenario DM particles never reach thermal equilibrium with the SM, due to the very small coupling between the two sectors, and hence they can not be produced via the standard WIMP scenario.
First DM particles could be produced by reheating and preheating mechanisms, but also by the FIMP mechanism.
However here we focus on a scenario where the bulk of the original DM particles were produced by the (p)reheating dynamics.
In fact, FIMP mechanism typically requires a portal coupling $\lambda_{\chi H}\sim 10^{-10}$ in order to reproduce the observed DM abundance.
In our case however, we have $\lambda_{\chi H}\ll 10^{-10}$, so that the FIMP production via SM particle annihilation and Higgs decays is subdominant with respect to the (p)reheating production.

Moreover, due to the dynamics within the dark sector, characterized by sizable couplings between DM particles, DM self-interactions could play a crucial role in the generation of the DM relic abundance.
In fact, one can consider a framework where the freeze-out proceeds via $N$-to-$n$ number-changing processes, where $N$ DM particles annihilate into $n$ of them (with $N>n\geq 2$).
This possibility was first studied in Ref.~\cite{Carlson:1992fn} and recently named the `SIMP paradigm'~\cite{Hochberg:2014dra}.
This has been entertained recently in strongly self-interacting DM models that annihilate in number-depleting 3-to-2~\cite{Dolgov:1980uu,Carlson:1992fn,Hochberg:2014dra,Hochberg:2014kqa,Bernal:2015bla,Bernal:2015lbl,Lee:2015gsa,Choi:2015bya,Hansen:2015yaa,Bernal:2015ova,Kuflik:2015isi,Hochberg:2015vrg,Choi:2016hid,Pappadopulo:2016pkp,Farina:2016llk,Choi:2016tkj,Dey:2016qgf,Choi:2017mkk,Ho:2017fte,Dolgov:2017ujf,Garcia-Cely:2017qpx,Choi:2017zww,Chu:2017msm,Duch:2017khv,Heikinheimo:2018esa} or 4-to-2 interactions~\cite{Bernal:2015xba,Bernal:2017mqb,Heikinheimo:2017ofk,Herms:2018ajr,Bernal:2018ins}.

In the case where the Higgs portal is very suppressed, the Boltzmann equation which describes the evolution of the DM number density $n(T')$ is:
\begin{equation}
	\frac{dn}{dt}+3\,H(T)\,n=-\langle\sigma v^2\rangle_{3\to 2}\left[n^3-n\,n_\text{eq}^2\right]-\langle\sigma v^3\rangle_{4\to 2}\left[n^4-n^2\,n_\text{eq}^2\right],
\end{equation}
where $H(T)$ is the Hubble expansion rate as a function of the temperature $T$ of the visible sector and $n_\text{eq}(T')$ represents the equilibrium DM number density at a dark temperature $T'$.
The factors $\langle\sigma v^2\rangle_{3\to 2}$ and $\langle\sigma v^3\rangle_{4\to 2}$ correspond to the generalized annihilation cross sections of the 3-to-2 and 4-to-2 DM annihilations, respectively.
In the case where the DM stability is guaranteed by a $\mathbb{Z}_2$ symmetry, 3-to-2 processes are forbidden and the DM annihilations are driven by the 4-to-2 processes, with a cross section that in the non-relativistic limit is given by~\cite{Bernal:2015xba}
\begin{equation}
	\langle\sigma v^3\rangle_{4\to 2}\sim\frac{27\sqrt{3}}{8\pi}\frac{\lambda_\chi^4}{m_\chi^8}.
\end{equation}

Let us recall that the entropies of the dark and the visible sectors are separately conserved, because they were always kinetically decoupled from each other.
It is then useful to compute the entropy ratio $\xi$ between them, which is also a conserved quantity.
In particular, if the freeze-out happens non-relativistically, one has
\begin{equation}
\label{entropy:cor}
\xi \equiv \frac{s}{s'}  \sim \frac{T'_\text{FO}}{3.6~\text{eV}\,(1+ 2.5\,T'_\text{FO}/m_\chi) \times  \Omega_\chi h^2}\,,
\end{equation}
where $s$ and $s'$ are the entropy densities of the SM sector and the dark sector, respectively, and $T'_\text{FO}$ is the temperature of the dark sector when the DM freeze-out takes place~\cite{Carlson:1992fn,Bernal:2015xba}.
Substituting $\Omega_\chi h^2 \sim 0.12$~\cite{Aghanim:2018eyx} and the solution of $T'_\text{FO}$~\cite{Bernal:2015xba}, the proper DM abundance requires 
\begin{equation}
\label{entropy:observed}
\xi  \sim  7\times 10^9 \, \left(\frac{m_\chi}{\text{GeV}}\right) \frac{1}{1 + 0.03 \left(\lambda_\chi\,\frac{\text{GeV}}{m_\chi}\right)^{4/7}}\,.
\end{equation}
The entropy ratio is nearly proportional to the DM mass, up to a weak dependence on the self-coupling $\lambda_\chi$. Moreover, to match the observed DM relic abundance there is always a $\xi\gg 10^2$ for each DM mass. The model therefore never leads to observable extra radiation~\cite{Carlson:1992fn}, consistent with BBN/CMB bounds.

\begin{figure}[t!]
        \begin{center}
        \includegraphics[width=.49\textwidth]{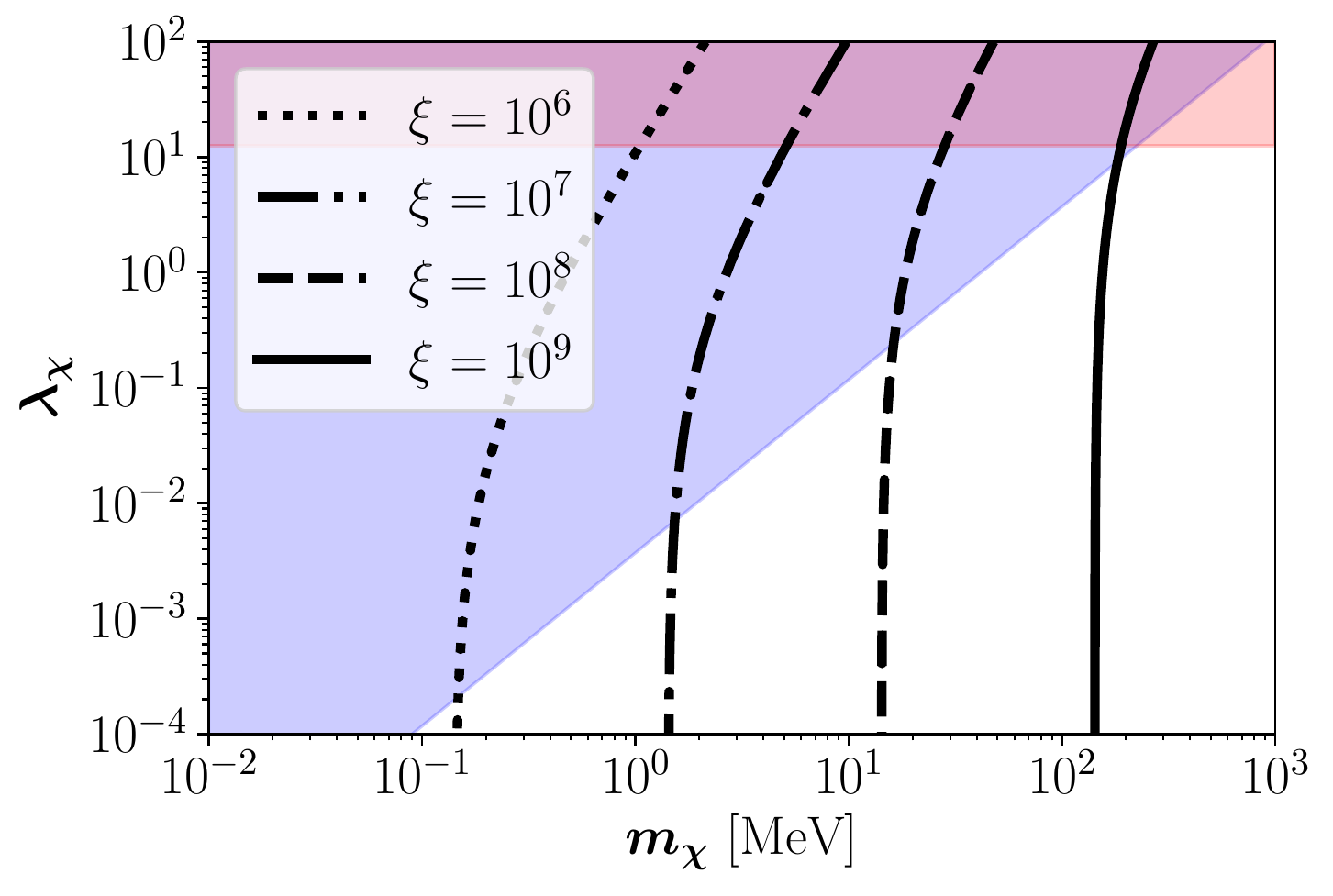}
        \includegraphics[width=.49\textwidth]{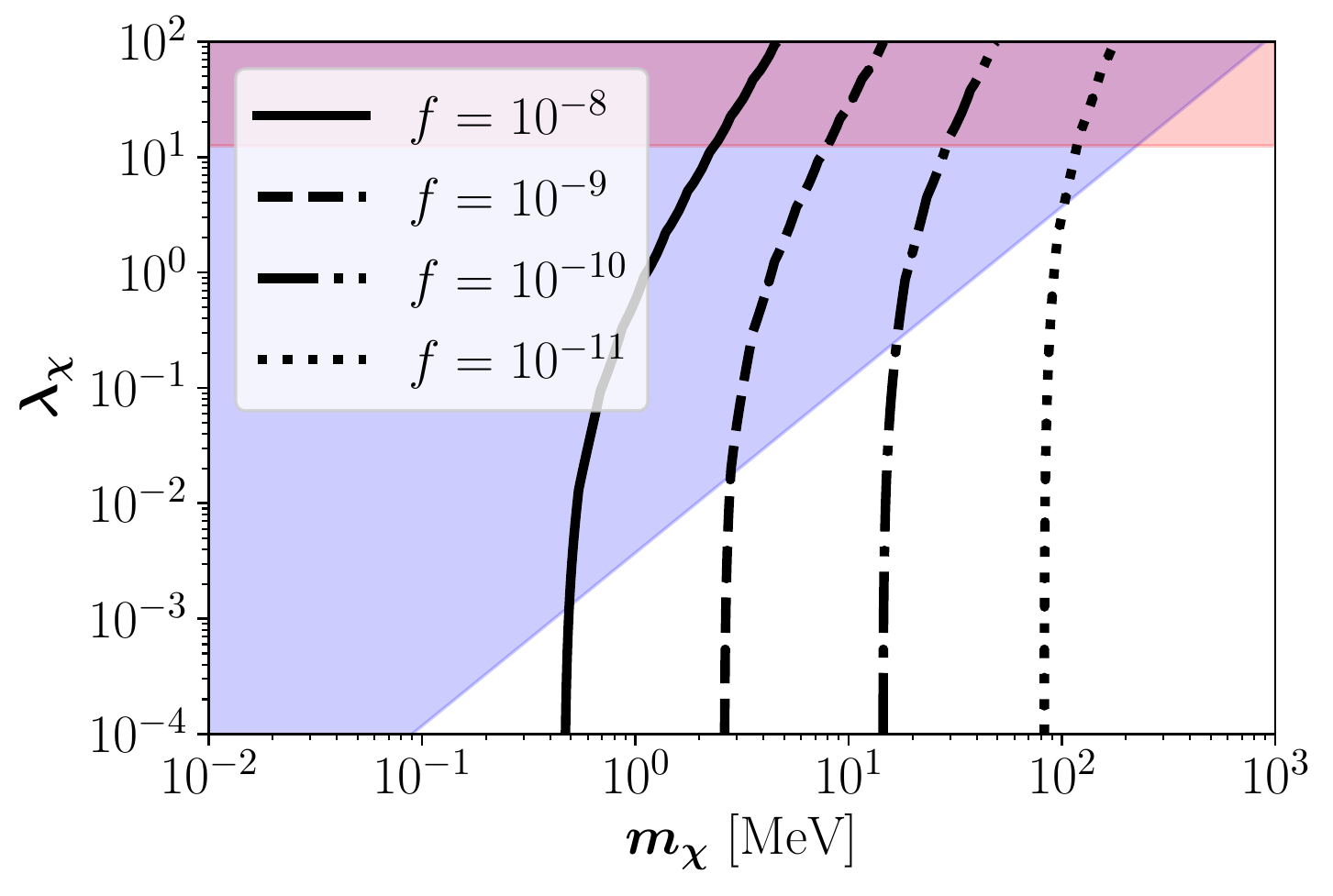}
		\caption{Values of $\xi$ and $f$ that yield the observed DM relic density, in the ($m_\chi$, $\lambda_\chi$) plane.
		The light blue region corresponds to $\sigma/m_\chi > 1.25$~cm$^2$/g and it is excluded by cluster observations.
		The area where $\lambda_\chi > 4\pi$ is shown in light red.
        }\label{fig:cannibal}
        \end{center}
\end{figure}

Fig.~\ref{fig:cannibal} shows the entropy ratio $\xi$ (left panel) and the energy density ratio $f$ (right panel) between the two sectors needed in order to reproduce the observed DM abundance, in the ($m_\chi$, $\lambda_\chi$) plane.
Let us emphasize that even if the entropy ratio is a constant, the energy density ratio in general is not.
For a fixed number of relativistic degrees of freedom, $f$ does not vary while the DM stays relativistic.
During that period, the two quantities are related by
\begin{equation}
	f=\frac{g_{\star S}^{4/3}}{g_\star}\,\xi^{-4/3}.
\end{equation}
In Fig.~\ref{fig:cannibal} the upper left corner (light blue) shows the region where DM elastic scatterings are too strong, and in tension with the Bullet Cluster constraints: $\sigma/m_\chi < 1.25$~cm$^2$/g at 68\% CL~\cite{Markevitch:2003at,Clowe:2003tk,Randall:2007ph}.
Additionally, the upper band corresponding to non-perturbative quartic couplings ($\lambda_\chi>4\pi$) is also discarded.

\section{Conclusion}
\label{conclusion}
In this work we have explored the possibility of reproducing the measured relic DM abundance from (p)reheating.
This scenario is interesting because it does not rely on interaction of DM particles with the SM particles, as required for standard freeze-out mechanism, of which no hint has been observed in direct and indirect searches.
While the production of DM during reheating (perturbative decay) is rather simple to estimate and the abundances can directly be obtained from the branching ratios of SM and DM from inflaton, the dynamics of preheating dynamics can not be solved analytically as the equation of evolution becomes non-linear after the initial stage of preheating.
We have used publicly available code {\tt LATTICEEASY} to simulate the dynamics of preheating for quadratic inflation and find that DM self-interactions can suppress the production of DM from preheating.
However, this suppression remains inadequate to reproduce the observed DM relic abundance, respecting the Bullet cluster bound.
We show that, in order to achieve further depletion in the DM number density, a non-relativistic phase of the inflaton before it decays completely to SM and a cannibalization mechanism needs to be invoked.
Note that the same DM quartic parameter which suppresses the preheating production, naturally generates cannibalization processes through 4-to-2 annihilations which can deplete the DM relic abundance.
In more general models, DM annihilation into other dark sector particles could help depleting the DM abundance, relaxing the length of the non-relativistic phase of the inflaton.

%\section{Conclusion}
%\section{Acknowledgements}
%\section{Bibliography}
%\input{Bibliography}

\section*{Acknowledgments}
The authors would like to thank Juan Pablo Beltrán Almeida and Supratik Pal for fruitful discussions.
We acknowledge partial support from the European Union Horizon 2020
research and innovation programme under the Marie Sk{\l}odowska-Curie: RISE
InvisiblesPlus (grant agreement No 690575)  and
the ITN Elusives (grant agreement No 674896).
NB is also supported by the Universidad Antonio Nariño grants 2017239 and 2018204, and 
by the Spanish MINECO under Grant FPA2017-84543-P. AC acknowledges support from 
Department of Science and Technology, India, through INSPIRE faculty fellowship, 
(grant no: IFA 15 PH-130, DST/INSPIRE/04/2015/000110). AP  thanks the computational facilities of Indian Statistical Institute, Kolkata. AP is supported by Council of Scientific and Industrial Research (CSIR), India, File no. 09/093(0169)/2015 EMR-I.
In addition to the software packages cited above, this research made use of IPython~\cite{Perez:2007emg}, Matplotlib~\cite{Hunter:2007ouj} and SciPy~\cite{SciPy}.

\bibliography{biblio}

\end{document}